\documentclass[a4paper,11pt]{article}
\usepackage{jinstpub}
\usepackage{comment}
\usepackage{lineno}
\usepackage[LGRgreek]{mathastext}
\usepackage{xcolor,soul}
\usepackage{arabtex} %to add the arabic words
\usepackage{utf8} %to add the arabic words
\setcode{utf8} %to add the arabic words

\usepackage{array}
\newcolumntype{P}[1]{>{\centering\arraybackslash}p{#1}}
\newcolumntype{M}[1]{>{\centering\arraybackslash}m{#1}}

\newcommand{\cb}{CeBr$_3$(LB)}

\title{RAAD: LIGHT-1 CubeSat's Payload for the Detection of Terrestrial Gamma-Ray Flashes}

\author[1,a,c]{A. Di Giovanni \note{Corresponding Author A. Di Giovanni: adriano.digiovanni@gssi.it}}
\author[b,c]{F. Arneodo\note{Corresponding Author F. Arneodo: francesco.arneodo@nyu.edu}}
\author[d]{A. Al Qasim}
\author[e]{H. Alblooshi}
\author[b]{F. AlKhouri}
\author[b,c]{L. Alkindi}
\author[d]{A. AlMannei}
\author[f]{M. L. Benabderrahmane}
\author[b,c]{G. Bruno}
\author[g]{V. Conicella}
\author[b]{O. Fawwaz}
\author[h]{G. Franchi}
\author[b]{S. Kalos}
\author[b,c]{P. Oikonomou}
\author[h]{L. Perillo}
\author[i,j]{C. Pittori}
\author[k]{M. S. Roberts}
\author[l]{R. Torres}

\affiliation[a]{Gran Sasso Science Institute,Viale Crispi n.7, 67100, L'Aquila, Italy}
\affiliation[b]{New York University Abu Dhabi, Saadiyat Island, Abu Dhabi, UAE}
\affiliation[c]{Center for Astro Particle, and Planetary Physics, New York University Abu Dhabi, Saadiyat Island, Abu Dhabi, UAE}
\affiliation[d]{UCL Mullard Space Science Laboratory, Holmbury Hill Road, Dorking, Surrey, RH5 6NT, UK}
\affiliation[e]{UAE Space Agency, Masdar City, Abu Dhabi, United
Arab Emirates}
\affiliation[f]{Mittelstra{\ss}e 3A,40721 Hilden, Germany}
\affiliation[g]{Gran Sasso Tech foundation, Viale Crispi n.7, 67100, L'Aquila, Italy}
\affiliation[h]{AGE Scientific srl, Capezzano Pianore, Lucca, Italy}
\affiliation[i]{ASI Space Science Data Center (ASI-SSDC), Via del Politecnico, Rome, Italy}
\affiliation[k]{Eureka Scientific, Oakland, Ca., USA}
\affiliation[j]{INAF, Osservatorio Astronomico di Roma (INAF-OAR), Via di Frascati, Monte Porzio Catone, Rome, Italy}
\affiliation[l]{University of Florence, Firenze, Italy}

\emailAdd{adriano.digiovanni@gssi.it; francesco.arneodo@nyu.edu}

\abstract{
The Rapid Acquisition Atmospheric Detector (RAAD), onboard the LIGHT-1 3U CubeSat, detects photons between hard X-rays  and soft gamma-rays, in order to identify and characterize Terrestrial Gamma Ray Flashes (TGFs). Three detector configurations are tested, making use of Cerium Bromide and Lanthanum BromoChloride scintillating crystals coupled to photomultiplier tubes or Multi-Pixel Photon Counters, in order to identify the optimal combination for TGF detection. High timing resolution, a short trigger window, and the short decay time of its electronics allow RAAD to perform accurate measurements of prompt, transient events. Here we describe the overview of the detection concept, the development of the front-end acquisition electronics, as well as the ground testing and simulation the payload underwent prior to its launch on December 21st, 2021. We further present an analysis of the detector's in-orbit system behavior and some preliminary results.
}

\arxivnumber{1234.56789} % Only if you have one

\begin{document}
\maketitle
\flushbottom

%%%%%%%%%%%%%%%%%%%%%%%%%%%%%%%%%%%%%%%%%%%%%%%%%%%%%%%%%%%%%%%%%%%%%%%%%%%%%%%%%%%
%% Introduction
%%%%%%%%%%%%%%%%%%%%%%%%%%%%%%%%%%%%%%%%%%%%%%%%%%%%%%%%%%%%%%%%%%%%%%%%%%%%%%%%%%%
\section{Introduction}
\label{sec:introduction}

% What are TGFs 
Terrestrial Gamma Ray Flashes (TGFs) are upward directed, highly luminous bursts of photons, with durations of less than 1 ms and energies from 10 keV reaching up to several tens of MeV\cite{Briggs2013,Dwyer2012,Fishman1994,Marisaldi2010,Nemiroff1997,Tavani2011}. They are produced in the high electric fields naturally occurring within thunderstorms, at altitudes of 10 - 15 km, when electrons are accelerated to relativistic speeds and produce photons through bremsstrahlung \cite{Dwyer20122,Dwyer2013,Khamitov2020,Nemiroff1997}. It is estimated that a sum of more than 400,000 TGFs are produced annually \cite{Briggs2013}, primarily concentrated around the equator \cite{Fishman1994}. 

%%%%%%%%%%%%%%%%%%%%%%%%%%%%%%%%%%%%%%%%%%%%%%%%%%%%%%%%%%%%%%%%%%%%%%%%%%%%%%%%%%%
%% Background
%%%%%%%%%%%%%%%%%%%%%%%%%%%%%%%%%%%%%%%%%%%%%%%%%%%%%%%%%%%%%%%%%%%%%%%%%%%%%%%%%%%

TGFs were first discovered by Fishman et al. using data obtained from the Burst and Transient Source Experiment (BATSE) on NASA's Compton Gamma Ray Observatory (CGRO) in 1994 \cite{Fishman1994}, and were found to be associated with lightning activity. BATSE was built to detect gamma-rays from celestial sources and thus was sensitive to energies of $\geq$20 keV. However, BATSE's sampling rate of $64\,$ms was longer than the typical sub-millisecond duration of a TGF, resulting in very few detections \cite{Fishman1994,Nemiroff1997}.

NASA's Reuven Ramaty High Energy Solar Spectroscopic Imager (RHESSI), originally built for studying solar flares, was able to detect TGFs as well. However, RHESSI's Germanium detectors were sensitive to photons with energies up to only 18 MeV. \cite{Smith2005}. Regardless, RHESSI's detections proved that TGFs can be detected by satellites operating il LEO (Low Earth Orbit) \cite{Cummer2005,Grefenstette2009}. The ``Astro-Rivelatore Gamma a Immagini Leggero'' (AGILE) and the Gamma-Ray Burst Monitor (GBM) onboard NASA's Fermi Gamma-ray Space Telescope were able to detect the full energy spectrum of TGFs \cite{Fuschino2009,Marisaldi2010,Marisaldi2011,Atwood2009,Meegan2009,Briggs2010,Roberts2018}. As the aforementioned missions were optimized to detect fainter phenomena of longer duration, their detectors suffered from pileup and saturation when observing high luminosity and energetic TGFs.

Thus far, TGFs have been primarily studied as by-products of missions designed to detect other phenomena, thus creating an opportunity for specialized missions. One such mission is European Space Agency's Atmosphere-Space Interactions Monitor (ASIM), installed in 2018 on the ISS, which managed to confirm the hypothesized TGF models originating from previous studies by combining observations between $50\,$keV - $30\,$MeV with optical data  \cite{Neubert2009,BjrgeEngeland2022}.

Even though TGFs' bright and brief nature made them hard to characterise with the non-specialised equipment of the initial missions, it also made their detection possible using detectors small enough to fit within CubeSats (i.e., pico-satellites developed at reduced costs). Thus we proposed the Rapid Acquisition Atmospheric Detector (RAAD), an instrument specifically designed for TGF detection, as the payload of the LIGHT-1 CubeSat mission. Table \ref{table:mission_comparison} shows the performance characteristics of LIGHT-1 compared to previous missions. The RAAD acronym was also chosen for its similarity to the Arabic word "Ra'ad" ( \<رعد> ) which means thunder. 

In this paper, we provide a detailed description of the payload electronics, ground tests and simulations, as well as some preliminary in-orbit data. A prototype of RAAD was built and tested at NYUAD, providing proof of concept. Details on the prototype can be found in \cite{DiGiovanni2019}. Details on the CubeSat's bus and subsystems can be found in  \cite{Almazrouei2021}.

\begin{table}[ht]
\centering
\begin{tabular}{ |M{2.1cm}||M{3cm}|M{2cm}|M{4cm}|M{2.5cm}| }
 \hline
    {\bf Mission} & {\bf Trigger Window (ms)} & {\bf Resolution ($\mu$s)} & {\bf Dead Time Per Event ($\mu$s)} & {\bf Sensitivity (MeV)}\\
  \hline
  \hline
    BATSE   &   64   &  2    &     6   & 0.02  $\div$ 2   \\
  \hline
    RHESSI  &  None  &  0.95 &     8   & 0.003 $\div$ 17 \\
  \hline
    AGILE   &  0.293 &  1    &   65    & 0.3   $\div$ 100 \\
  \hline
    FERMI   &   16   &  2    &   2.6   & 0.001 $\div$ 40 \\
  \hline
    LIGHT-1 &   0.5  &  0.5  &  0.04 & 0.02 $\div$ 3 \\
  \hline
  \hline
\end{tabular}

\caption{Characteristics of LIGHT-1 compared with previous missions: BATSE \cite{Fishman1994}, RHESSI \cite{Smith2002}, AGILE \cite{Labanti2009,Fuschino2009}, and FERMI \cite{Meegan2009}.}

\label{table:mission_comparison}
\end{table}

%%%%%%%%%%%%%%%%%%%%%%%%%%%%%%%%%%%%%%%%%%%%%%%%%%%%%%%%%%%%%%%%%%%%%%%%%%%%%%%%%%%
%% The LIGHT-1 Mission
%%%%%%%%%%%%%%%%%%%%%%%%%%%%%%%%%%%%%%%%%%%%%%%%%%%%%%%%%%%%%%%%%%%%%%%%%%%%%%%%%%%
\section{The LIGHT-1 Mission}
\label{sec:LIGHT-1-mission}

%%%%%%%%%%%%%%%%%%%%%%%%%%%%%%%%%%%%%%%%%%%%%%%%%%%%%%%%%%%%%%%%%%%%%%%%%%%%%%%%%%%
%% Concept

\subsection{Mission Concept}
\label{subsec:concept}

Absorption by the atmosphere can suppress TGFs, therefore, solely measuring the gamma-ray emission would not provide complete information about their characteristics \cite{Mailyan2019}. As a result, a combined measurement is, in general, preferred over a single detection channel. Combining atmospheric gamma-ray detection with X and gamma-ray surveys or lightning radar measurements could provide insightful information on the production mechanisms of TGFs \cite{Chronis2016,Mailyan2018}. Under this premise, a LEO operated gamma-ray detector, together with correlated ground and space observations and measurements, would be the ideal probe to pursue a TGF science program.

LIGHT-1 is a 3U CubeSat built primarily to study TGFs. Yet, LIGHT-1's mission extends to measuring orbital radiation and space-qualifying different technologies for the detection of prompt, highly energetic, and intense emissions typical of transient events. For size reference, one U of a CubeSat consists of a cube of about $\mathrm{10 ~cm ~\times~10 ~cm ~\times ~10 ~cm}$. A little less than two units of LIGHT-1 is dedicated to RAAD, while the rest is dedicated to the subsystems of the CubeSat (reaction wheels, onboard computer, attitude control system, electrical power system, etc.). A review on operating gamma-ray detectors onboard CubeSats is available in \cite{Arneodo2021}.

RAAD utilizes three types of detectors, by coupling two types of photosensors to two types of scintillating crystals. In fact, the LIGHT-1 mission is also meant to conduct a direct comparison between different detector configurations to find which is best suited for TGF detection. LIGHT-1 is the first mission to conduct a direct comparison between Cerium Bromide and Lanthanum BromoChloride scintillating crystals, coupled to either photomultiplier tubes or silicon photomultipliers.

%%%%%%%%%%%%%%%%%%%%%%%%%%%%%%%%%%%%%%%%%%%%%%%%%%%%%%%%%%%%%%%%%%%%%%%%%%%%%%%%%%%
%% Flight Schedule

\subsection{Mission Schedule}
\label{subsection:mission-schedule}

\begin{figure}[t]
    \centering
    \includegraphics[width=13cm]{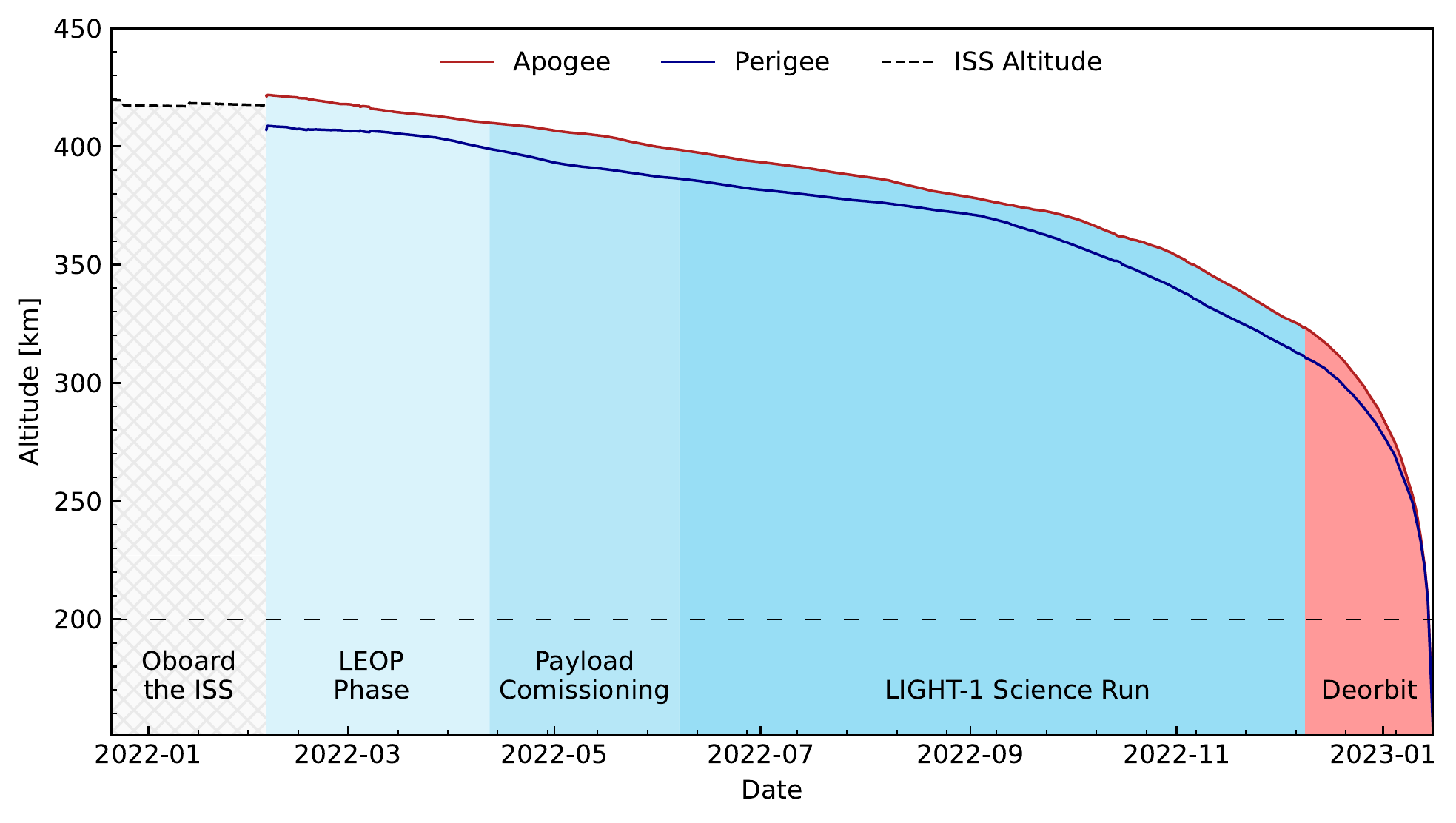}
    \caption{The life of the LIGHT-1 mission. Apogee and perigee altitudes of the satellite are shown in red and blue respectively. The dotted black line starting on the SpaceX Launch date (2021-12-21) until the deployment date (2022-02-03) is the altitude of the International Space Station where LIGHT-1 was stowed prior to the start of the mission. Different colors denote the different operation regimes: Launch and Early Orbit Phase (LEOP), Payload Commissioning, Science Run, and de-orbit.}
    \label{fig:light1life}
\end{figure}

The LIGHT-1 CubeSat was launched on the 21st of December, 2021, onboard a Space-X Falcon9/Dragon from the Kennedy Space Center, directed at the International Space Station (ISS). The handling of LIGHT-1 for the launch and subsequent deployment was taken care by the Japan Aerospace Exploration Agency (JAXA). Once on the ISS, it was deployed via the Japanese Experiment Module (JEM) on the 3rd of February, 2022, in a 51.6\textdegree orbit at an initial altitude of $\mathrm{408~km}$. 

Figure \ref{fig:light1life} shows the operation regimes of the LIGHT-1 mission as a function of time. During the Launch and Early Orbit Phase (LEOP), the satellite bus developed by NanoAvionics was fully tested and optimized for the science program operations. The payload was powered on for the completion of one full orbit on the 16th of March, 2022 to check its vital parameters. The commissioning of the payload began on the 6th of April, 2022. On the 25th of May, 2022 LIGHT-1 entered the Science Run mode. 
Communications with LIGHT-1 stopped on January 18th, 2023, marking the start of the de-orbiting phase and the end of the mission.

%%%%%%%%%%%%%%%%%%%%%%%%%%%%%%%%%%%%%%%%%%%%%%%%%%%%%%%%%%%%%%%%%%%%%%%%%%%%%%%%%%%
%% The Payload

\section{Scientific Payload Specifications}
\label{sec:payload-specifications}

\begin{table}[ht]
\centering
\begin{tabular}{ |M{6cm}||M{3cm}|M{3cm}|  }
 \hline
 {\bf Parameter} & {\bf Design value} & {\bf Actual Value}\\
 \hline
  \hline
 Mass [kg]  & 2.05 $\pm$ 0.07    &1.981 $\pm$ 0.001 \\
  \hline
 Average Power Consumption [W] &   $<$ 5.9 W  &  $<$ 4.8 W\\
  \hline
 Data Downlink [MB/24 h] & 40  & $\sim$ 40 \\
  \hline
 Duty Cycle [\%]    &50 & $\sim$ 50\\
  \hline
 Life Time [months]& $\sim$ 6  & $\sim$ 10\\
  \hline
%From funding to Launch [years]& 3  & 3   \\
%  \hline
 Temperature operative range [$^{\circ}\mathrm{C}$]& 0 $\div$ 45  & 10$\div$ 40\\
  \hline
 \hline
\end{tabular}
\caption{Payload specification design in comparison with the values measured on the flight model.}
\label{table:mission_constraints}
\end{table}

RAAD is designed to resolve events hundreds of ns apart, measure the energy deposited, and assign a timestamp for comparison with TGF and lightning catalogs generated by other experiments.

\subsection{Detector Structure}
\label{subsec:detector-structure}
 
\begin{figure}[!ht]
    \centering
    \includegraphics[clip,trim={0, 0cm, 0, 1cm},width=15cm]{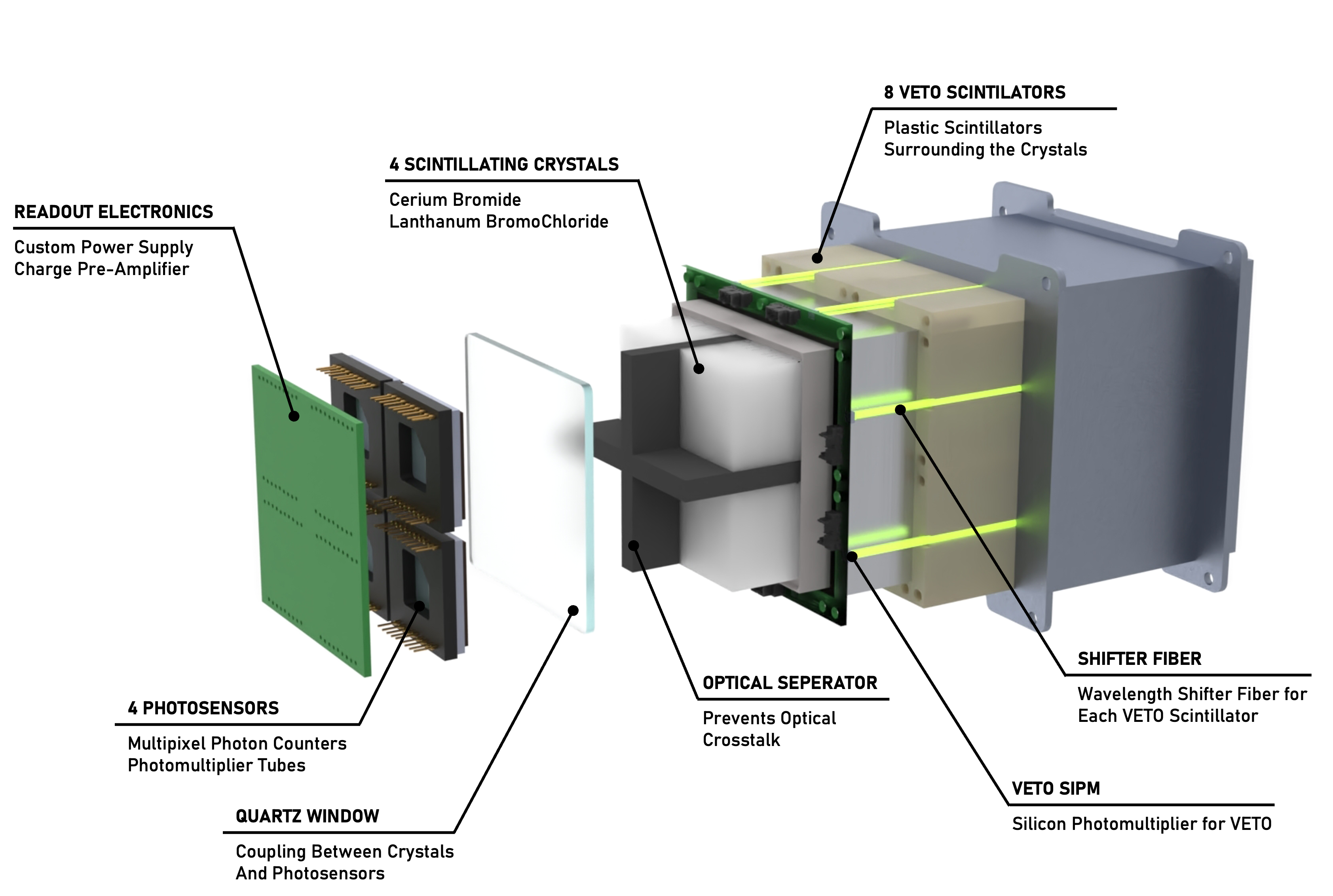} %clip,trim={0, 3cm, 0, 3cm},
    \caption{An explosion view of the MPPC CAD model. The PMT payload structure is identical except for the size of the photosensors.}
    \label{fig:3DLight1Payload}
\end{figure}

\begin{figure}[ht]
    \centering
    \includegraphics[width=8.5cm, angle=0]{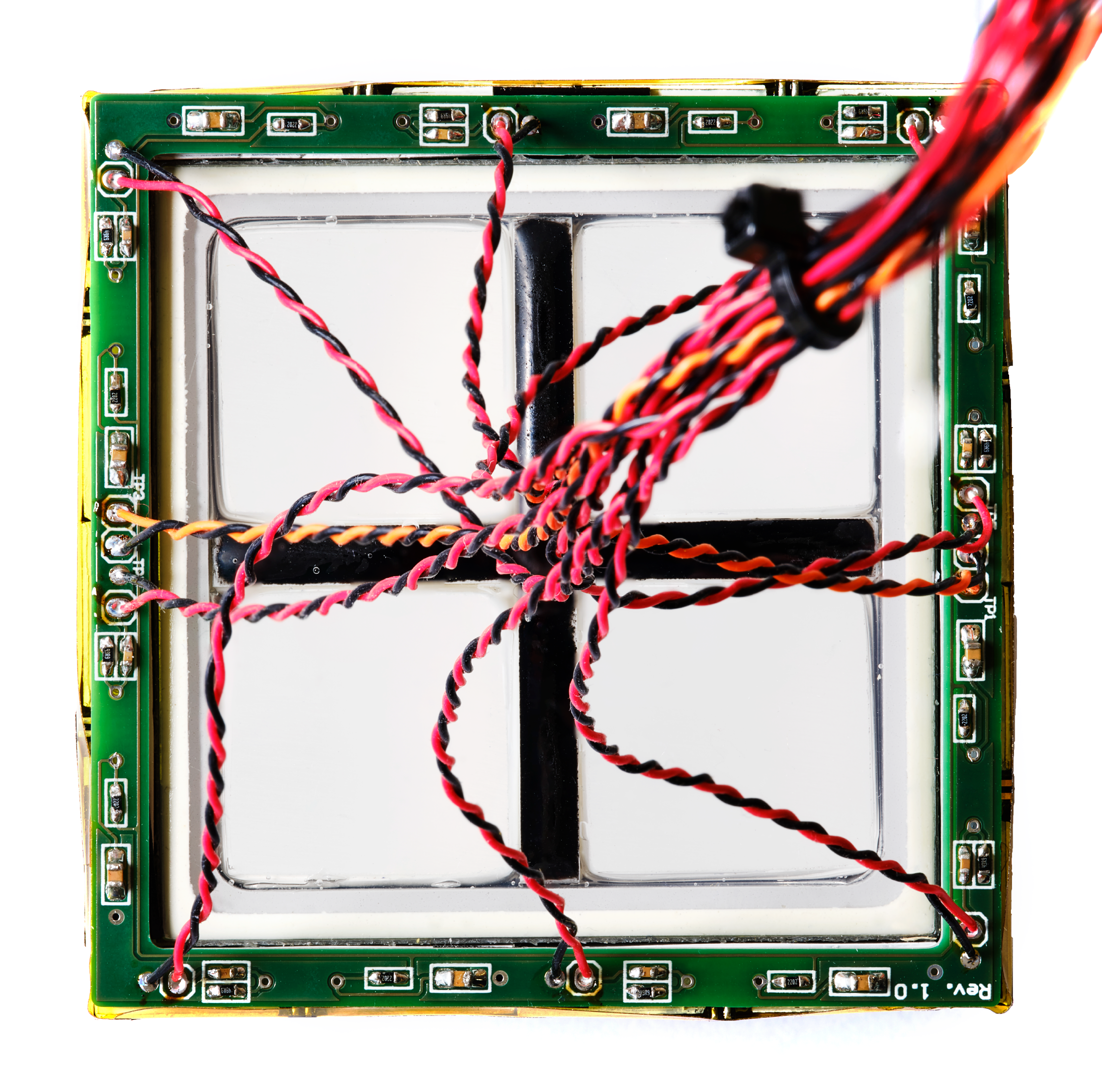}
    \caption{Top view of the \cb ~crystal array used in the MPPC payload, embedded in the VETO system and its readout electronics. The view of the PMT payload is identical. }
    \label{fig:crystal}
\end{figure}

RAAD consists of two detectors, different in size, fitting 1 U and 0.7 U of the spacecraft, respectively. The detector structure is shown in Figure~\ref{fig:3DLight1Payload}.
The smaller one, the {\it MPPC payload}, is equipped with four S13361-6050AE-04 Multi-Pixel Photon Counters (MPPCs) manufactured by Hamamatsu Photonics and coupled to a 4-channel Low Background Cerium Bromide (\cb) crystal array, manufactured by Scionix and shown in Figure~\ref{fig:crystal}. The larger detector, the {\it PMT payload}, is equipped with four R11265-200 photomultiplier tubes (by Hamamatsu Photonics) and coupled to one array of four scintillating crystals organized in one pair of \cb ~and one pair of Lanthanum BromoChloride (LBC), both manufactured by Scionix. We refer to each photosensor array coupled with the corresponding scintillating crystals as the detection target of RAAD. The differences between the two photosensors are shown in Table~\ref{table:photosensor-characteristics}.

\begin{table}[ht]
\centering
\begin{tabular}{ |M{8.5cm}||M{2cm}|M{3.5cm}|  }
 \hline
 {\bf Parameter} & {\bf R11265-200} & {\bf S13361-6050AE-04}\\
 \hline
  \hline
Type of photosensor &   PMT  & MPPC\\
  \hline
 Dimensions ($L\cdot D\cdot H$) [$\mathrm{mm^{3}}$]  & $26\cdot26\cdot19$    &$25\cdot25\cdot1.4$\\
  \hline
Mass [g] &   24  & 2\\
  \hline
 Max Sensitivity [nm] & 400  & 450 \\
  \hline
Quantum Efficiency - Photon Detection Efficiency [\%] &43 &  40\\
  \hline
 Operating Voltage [V]    &900 &  55\\
  \hline
Gain at working point & $\sim 10^6\div 10^7$  & $\sim 10^6\div 10^7$   \\
  \hline
Dark Counting Rate at working point, 300 K  [Hz]& < 1  & $> 10^7$ \\
  \hline
Operating Temperature [K] & 240 - 320  & 250 - 330 \\
    \hline
 \hline
\end{tabular}
\caption{Characteristics of the photosensors used in RAAD}
\label{table:photosensor-characteristics}
\end{table}

The performances of the crystals at different energies, shown in Table~\ref{table:crystal-characteristics}, make them complementary. The intrinsic activity of LBC, while on one hand an obvious nuisance, on the other hand, may provide an embedded calibration tool. The high dark counting rate of the MPPC does not represent a problem when triggering at $\mathrm{\gtrsim  20 \, keV}$ (see \ref{subsec:geant4}).

Each scintillating crystal unit is  $\mathrm{23~ mm ~\times ~23~ mm~ \times~ 45~ mm}$ in dimensions, with the whole scintillating crystal arrays of $\mathrm{60~ mm~ \times~ 60~ mm \times~ 48~ mm}$ in dimensions and $\mathrm{615~ g}$ and $\mathrm{595~ g}$ in mass, for the $\mathrm{CeBr_3(LB)}$ and LBC respectively. 

The four scintillating crystals are separated by Polytetrafluoroethylene (PTFE) to avoid optical cross-talk. In order to prevent the effects of water vapor contamination during the assembling phase, the hygroscopic crystal array is housed in an airtight container. Five sides are covered with aluminum, while on the sixth side, a 2 mm thick layer of fused silica is used to optically connect the photosensor array to the crystals.
 
 \begin{table}[ht]
\centering
\begin{tabular}{ |M{6cm}||M{3cm}|M{3cm}|  }
 \hline
 {\bf Parameter} & {\bf \cb} & {\bf LBC}\\
 \hline
  \hline
 Density [$\mathrm{g\cdot cm^{-3}}$]  & 5.1    &4.9\\
  \hline
 Hygroscopic &   Yes  & Yes\\
  \hline
 Light Emission Peak [nm] & 370  & 380 \\
  \hline
Energy Resolution at 122 keV [\%] &10 &  7\\
  \hline
Energy Resolution at 662 keV [\%]    &4 &  3\\
  \hline
Decay Time [ns]& 20  & 35   \\
  \hline
Intrinsic Activity  [$\mathrm{Bq \cdot cm^{-3}}$]& $<0.01$  & $\sim$ 1\\
  \hline
 \hline
\end{tabular}
\caption{Characteristics of the scintillating crystals used in RAAD.}
\label{table:crystal-characteristics}
\end{table}

\begin{figure}[ht]
    \centering
    \includegraphics[width=14cm, angle=0]{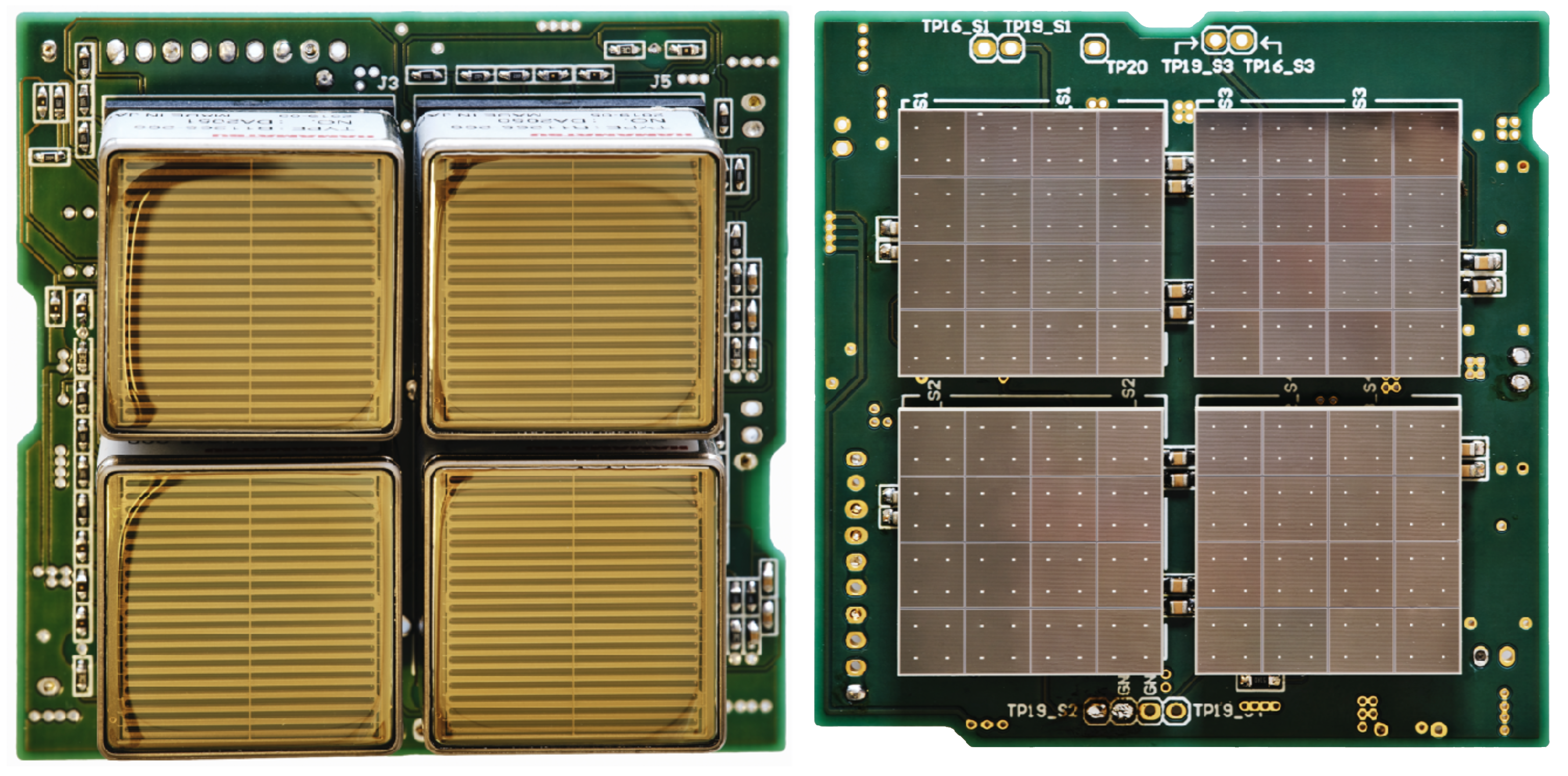}
    \caption{The two photosensor arrays from the payload of the LIGHT-1 mission. (left) The four R11265-200 photomultiplier tubes used in the PMT payload. (right) The four S13361-6050AE-04 Multi-Pixel Photon Counters used in the MPPC payload. Both sensors have been manufactured by Hamamatsu Photonics.}
    \label{fig:photosensorarrays}
\end{figure}

The front-end electronics are directly coupled to the back of the photosensor array as seen in Figure~\ref{fig:photosensorarrays}.

A detailed study and characterization of the detection concept is reported in \cite{DiGiovanni2019}.

A VETO system surrounds the detection target on four sides. Its purpose is to partially identify and suppress the charged particle induced background. It consists of eight independent units, each composed of one 5 mm thick plastic scintillator tile with an embedded wavelength shifter fiber and read out at one edge by a SMD Silicon photomultiplier (SiPM) manufactured by AdvanSiD srl (ASD-NUV1C-P-40). The CAD model of the VETO system is shown in Figure~\ref{fig:3DLight1Payload}.

\begin{figure}
    \centering
    \includegraphics[width=10cm]{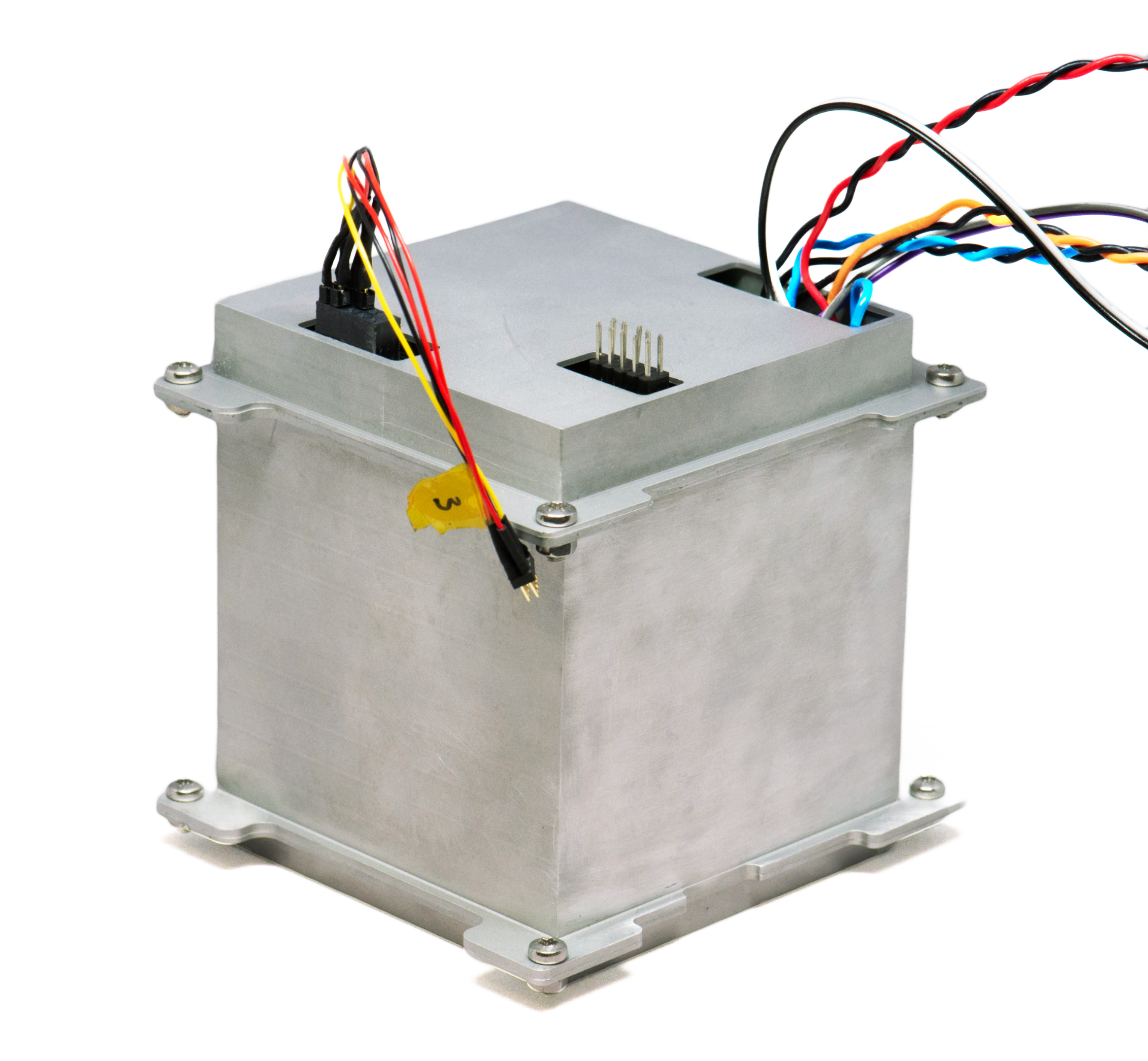}
    \caption{The fully assembled PMT payload. The view of the MPPC payload is identical. }
    \label{fig:modelassembled}
\end{figure}

The detection target along with the photosensor array and related electronics are placed inside an aluminum enclosure designed to be mechanically coupled to the CubeSat spacecraft. The final assembly of the PMT payload is shown in Figure~\ref{fig:modelassembled}. 

The effect of the aluminum walls has been evaluated by a Geant4 simulation, fully presented in Section~\ref{subsec:geant4}, indicating a hardware detection threshold for gamma rays of about 20 keV. 

An aerospace-grade silicon resin (Momentive RTV615) was used for filling the voids in the assembly in order to mitigate the effect of vibration on the structure and to provide effective electrical insulation for the high voltage ($\mathrm{-800~V\div-700~V}$) required to operate the payload. Overall, the multiple enclosing layers comply with the safety requirements of the mission, mitigating the risk of fragmentation.

%%%%%%%%%%%%%%%%%%%%%%%%%%%%%%%%%%%%%%%%%%%%%%%%%%%%%%%%%%%%%%%%%%%%%%%%%%%%%%%%%%%
%% Redout Electronics

\subsection{Payload Readout and Control Electronics}
\label{subsec:payload-electronics}

RAAD electronics consist of three main components. 
\begin{enumerate}
    \item VETO readout electronics
    \item Photosensor power supply
    \item Front-end and Controller boards
\end{enumerate}

\begin{figure}[ht]
    \centering
    \includegraphics[width=14cm, angle=0]{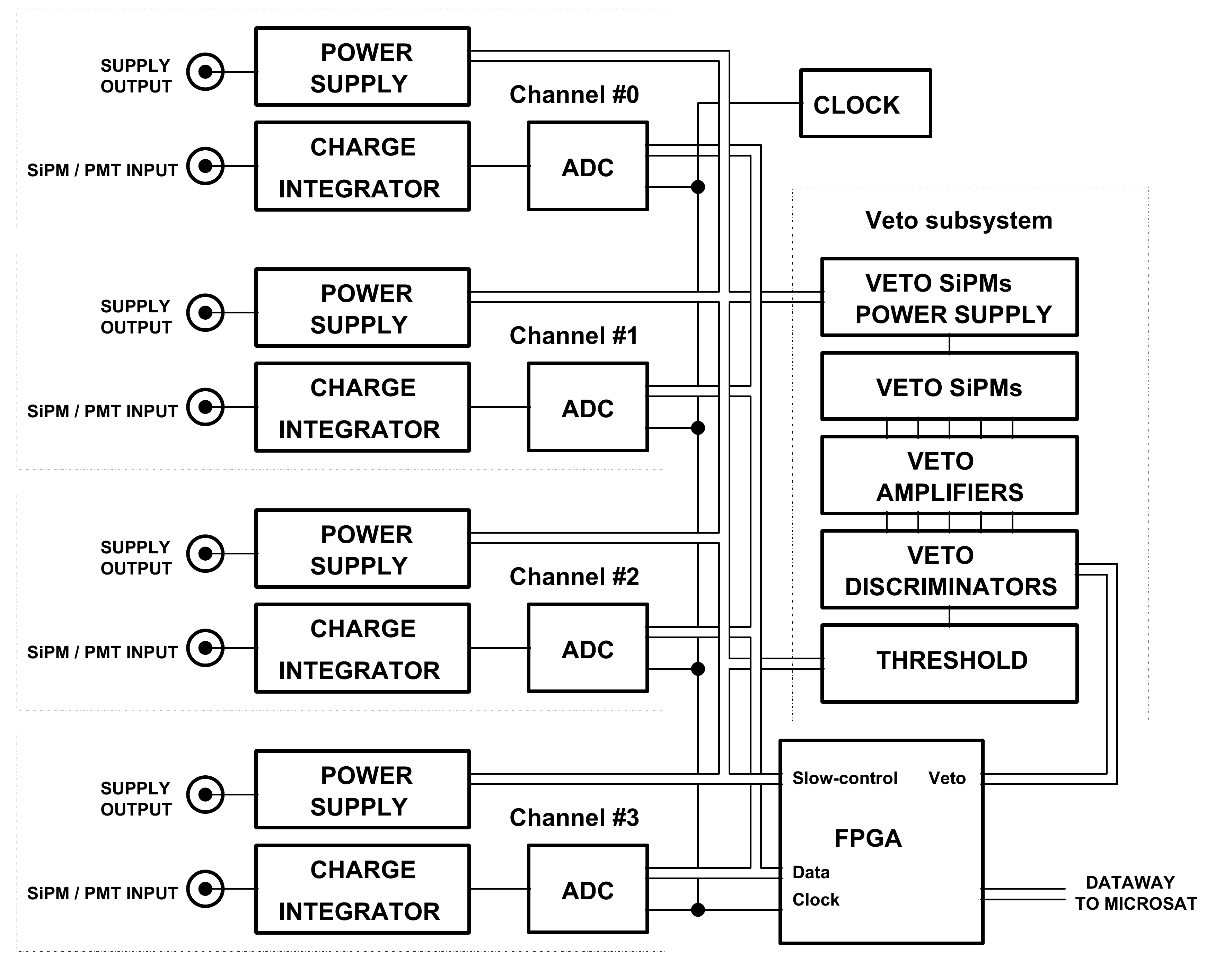}
    \caption{Block diagram of the RAAD electronics}
    \label{fig:block-diagram}
\end{figure}

A block diagram for the electronics is shown in Figure~\ref{fig:block-diagram} and its main characteristics are summarized in Table~\ref{table:payload_ele_characteritics}.

\begin{table}[ht]
\centering
\begin{tabular}{ |M{6cm}||M{6cm}||  }
 \hline
 {\bf Parameter} & {\bf Design value} \\
 \hline
  \hline
 Average Power Consumption [W] &   $<$ 5.9  \\
  \hline
 Mass [g] & 58 (MPPC), 66 (PMT) \\
  \hline
ADC resolution range [bits]   & 10 $\div$ 16\\
  \hline
 Components & SMD, all COTS, Automotive grade \\
  \hline
 Operating modes & noise, default, science, custom, safe\\
  \hline
 \hline
\end{tabular}
\caption{The main design characteristics of the RAAD electronics}
\label{table:payload_ele_characteritics}
\end{table}

\begin{figure}[ht]
    \centering
    \includegraphics[width=13cm, angle=0]{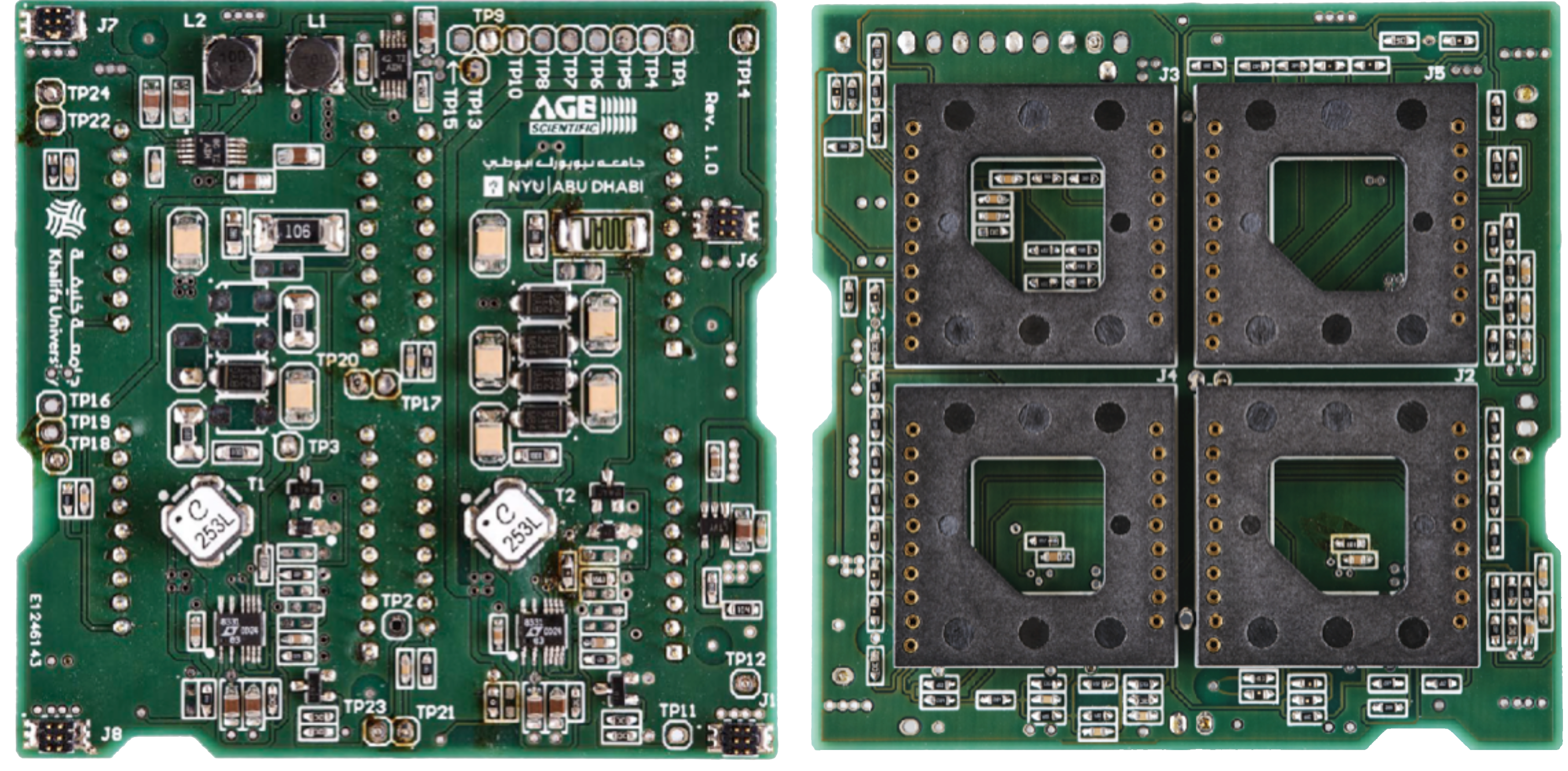}
    \caption{The custom made power supply electronics of the LIGHT-1 PMT payload which is used to bias the photomultiplier tubes at variable gains, depending on the science case.}
    \label{fig:electronics}
\end{figure}

The power supply electronics of the MPPC and PMT payloads are shown in Figure~\ref{fig:electronics}.

With the single exception of the photosensor sockets, no connectors have been used in the LIGHT-1 payload to maximize compactness and reliability.

\subsection{Firmware Overview}
\label{subsec:firmware-overview}

The onboard firmware is designed in order to assign all data collected to specific categories (i.e. gamma and charged particle-induced events, TGF events, and orbit monitoring data). 

\begin{figure}
    \centering
    \includegraphics[width=13cm]{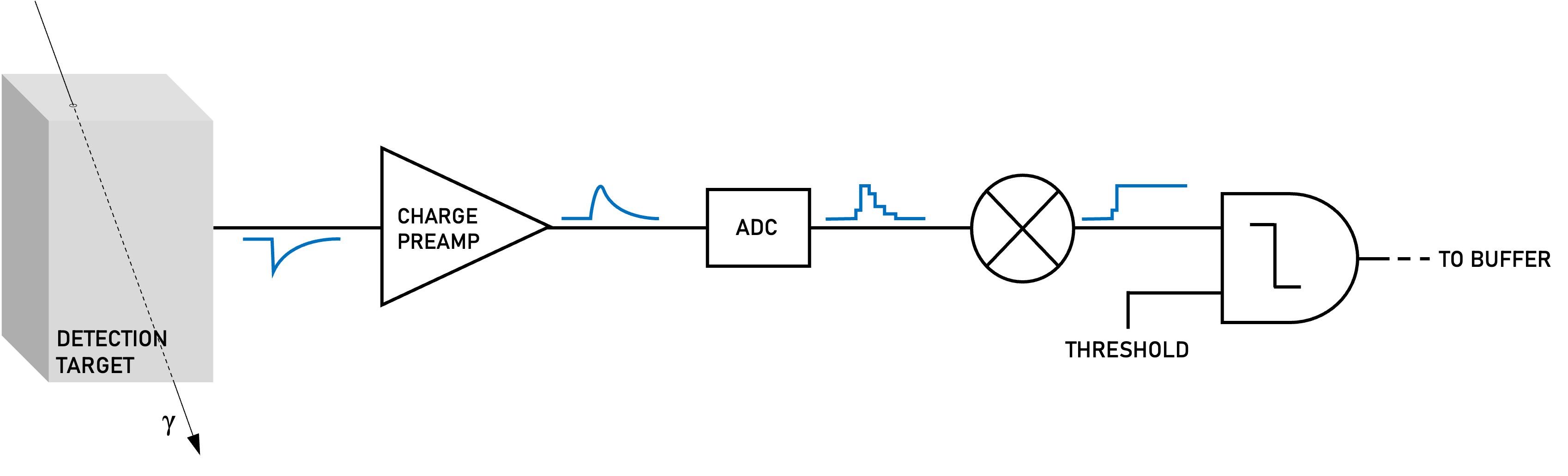}
    \caption{A schematic description of the detection pipeline of LIGHT-1. A particle interaction triggers a signal on the photosensor that is then amplified and digitized. If the signal is above a software set threshold it is recorded and sorted to the corresponding buffer.}
    \label{fig:detection-pipeline}
\end{figure}

The detection strategy can be summarised in a series of steps listed below ( see Figure~\ref{fig:detection-pipeline}).
\begin{enumerate}
    \item Particle interaction and signal generation in the detector;
    \item the charge amplifier integrates the signal. Its output is sampled by an ADC (16 bits, 10 MHz) and passed through a matched filter \cite{Arneodo2021} to extract the peak charge value;
    \item if the charge value is larger than the detection threshold, an event timestamp is assigned and the event is sent to the spacecraft bus.
    \end{enumerate}

The payload data generated are organized into four buffers, each designed to store different types of events as shown in Table~\ref{table:payload_BUFFER_characteristics}.

\begin{table}[ht]
\centering
\begin{tabular}{ |M{2cm}||p{10cm}||  }
 \hline
 {\bf Data Buffer} & {\bf Description} \\
 \hline
  \hline
NonVeto  &   All the events in which there is no VETO flag. The ADC resolution is lowered to 10 bits, timestamp at 1 ms resolution \\
  \hline
Veto & All the events in which there is at least one (out of eight) VETO flag active. The ADC resolution is lowered to 14 bits, timestamp at 100 $\mu s$ resolution \\
  \hline
TGF & All the events falling into a coincidence window of 500 $\mu s$ in which there is no VETO flag. The ADC resolution is nominal (16 bits), 500 ns timestamp resolution\\
  \hline
 Orbit & Payload housekeeping events collected once every 20 s including the payload temperature, the particle rate of each channel and the VETO, the detector operating voltage, the payload operating configuration (e.g. noise, default, flight mode), and all the working parameters required to run diagnostic checks\\
  \hline
 \hline
\end{tabular}
\caption{Data buffer protocol used for the scientific data of the LIGHT-1 mission.}
\label{table:payload_BUFFER_characteristics}
\end{table}

The science data is transmitted to the ground using an S-Band system during daily passes above ground stations dedicated to the mission \cite{Almazrouei2021}. Each contact (up 6 times per day) can last up to 10 minutes. Spacecraft telemetry, which includes housekeeping data, operational commands, and payload scripts, is transmitted using a UHF transceiver.

 %%%%%%%%%%%%%%%%%%%%%%%%%%%%%%%%%%%%%%%%%%%%%%%%%%%%%%%%%%%%%%%%%%%%%%%%%%%%%%%%%%%
%% Simulations and Tests
%%%%%%%%%%%%%%%%%%%%%%%%%%%%%%%%%%%%%%%%%%%%%%%%%%%%%%%%%%%%%%%%%%%%%%%%%%%%%%%%%%%

\section{Pre-Flight Simulations and Tests}
\label{sec:simulations-tests}

In order to space qualify RAAD, all the payload components and the payload as a whole have been subjected to several environmental and functionality tests, in order to assess the surviving capability to the various regimes typical of the mission (launch phase, thermal stress, radiation, etc ...).

\subsection{Stress Analysis}
\label{subsec:stress-analaysis}

Prior to conducting vibrations tests on the physical assembly of the payload, a stress analysis was conducted to study the effect of the launch conditions on the structure. In particular, a Finite Element Analysis was conducted in order to calculate the frequency and peak amplitude of the first 20 vibrational modes. 

 The large computational requirement for such analyses forced a geometric simplification of the 3D models used. Yet the simplification was within a reasonable range so as to not contribute more than 10\% of the payload's total mass.

 The general requirement for space qualifying such equipment to survive launch is for the natural modes to be above 100 Hz (see \cite{esa-testing}). Through the simulation, we verified that our design fulfills this requirement as all normal modes are between 1200 Hz and 2000 Hz.

% \begin{figure}[ht]
%     \centering
%     \includegraphics[width=14cm, angle=0]{media/StressAnalysisSimulation.png}
%     \caption{External deformation of the MPPC payload of LIGHT-1 under the first two normal modes at 1.2 kHz and 1.9 kHz respectively. The top and bottom rows show the two extremal modes for the first and second normal modes respectively}
%     \label{fig:StressSimulations}
% \end{figure}

%%%%%%%%%%%%%%%%%%%%%%%%%%%%%%%%%%%%%%%%%%%%%%%%%%%%%%%%%%%%%%%%%%%%%%%%%%%%%%%%%%%
%% Geant4 Simulations

\subsection{Geant4 Simulations}
\label{subsec:geant4}
Using the Geant4 framework a C++ particle physics simulation was developed to evaluate the efficiency of the detectors, as well as the background energy deposition on the crystals and the veto caused by trapped charged particles in the earth's magnetic field. The detector geometry was simplified to increase performance. The simulation code and analysis can be found in \cite{Geant2022}.

\begin{figure}
    \centering
    \includegraphics[width=13cm]{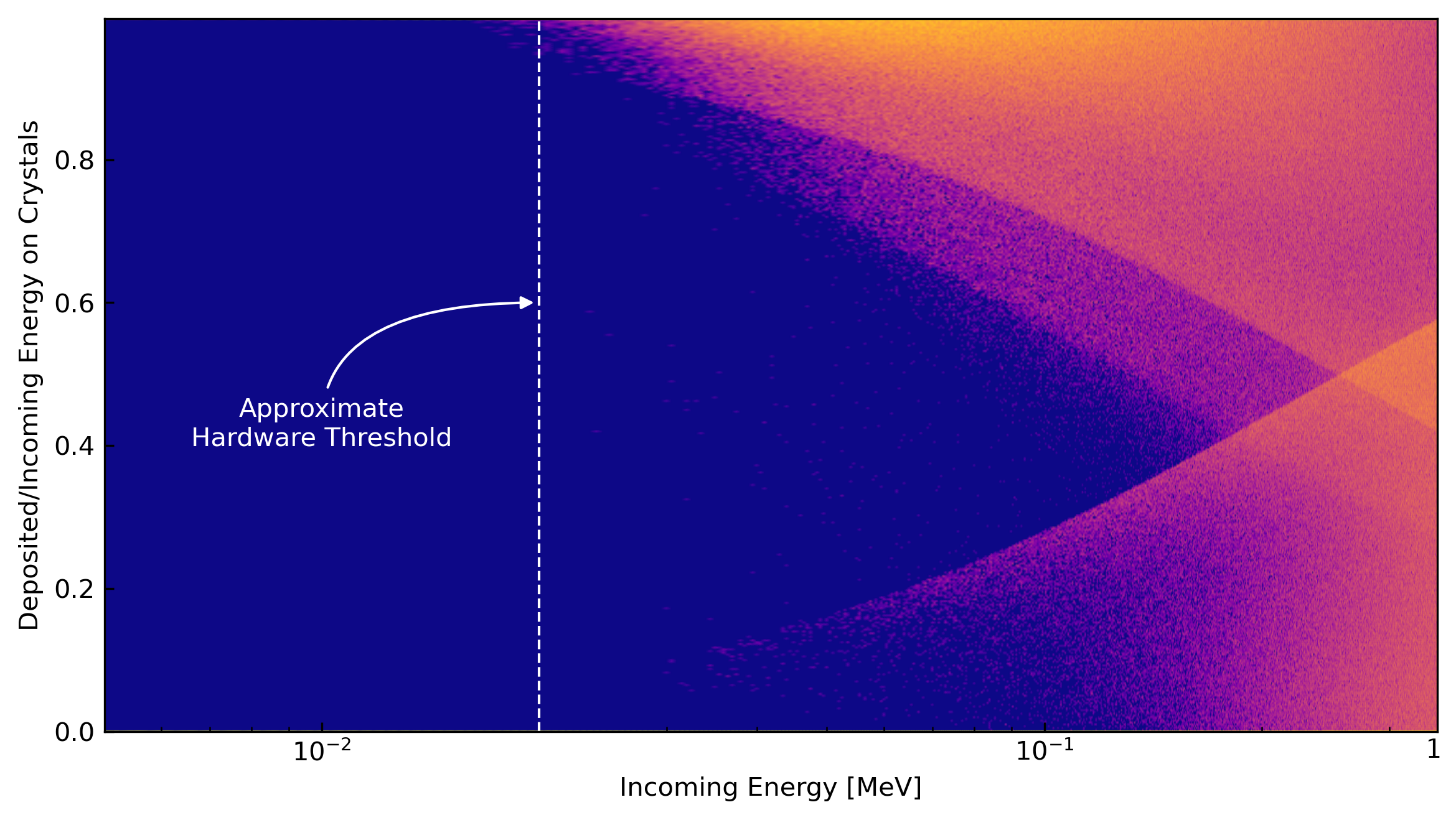}
    \caption{Geant4 hardware threshold estimation. Single photons were sent to the detector with equiprobable energy ranging from $1\ keV$ to $1\ MeV$. A histogram of the ratio of energy deposited on the crystals to incoming particle energy is plotted as a function of incoming energy. Brighter colors correspond to higher probability of observing the ratio. For example, most particles with energy 100 keV deposit around 90\% of their energy in the crystals, while almost none of them deposit between 30\% and 60\%. We can conclude that particles start depositing energy at $20\ keV$ and that they deposit roughly 100\% of their energy in the crystals.}
    \label{fig:geant4-hardware-threshold}
\end{figure}

We were able to estimate the hardware threshold of the detector by sending photons with energies ranging from $1\ keV$ to $1\ MeV$ and recording the energy deposited on each component of the detector. Fig.~\ref{fig:geant4-hardware-threshold} shows the ratio of the energy deposited to the crystals over the incoming energy as a function of the incoming photon energy. We see that we start detecting photons at roughly $20\ keV$.  

\begin{figure}
    \centering
    \includegraphics[width=13cm]{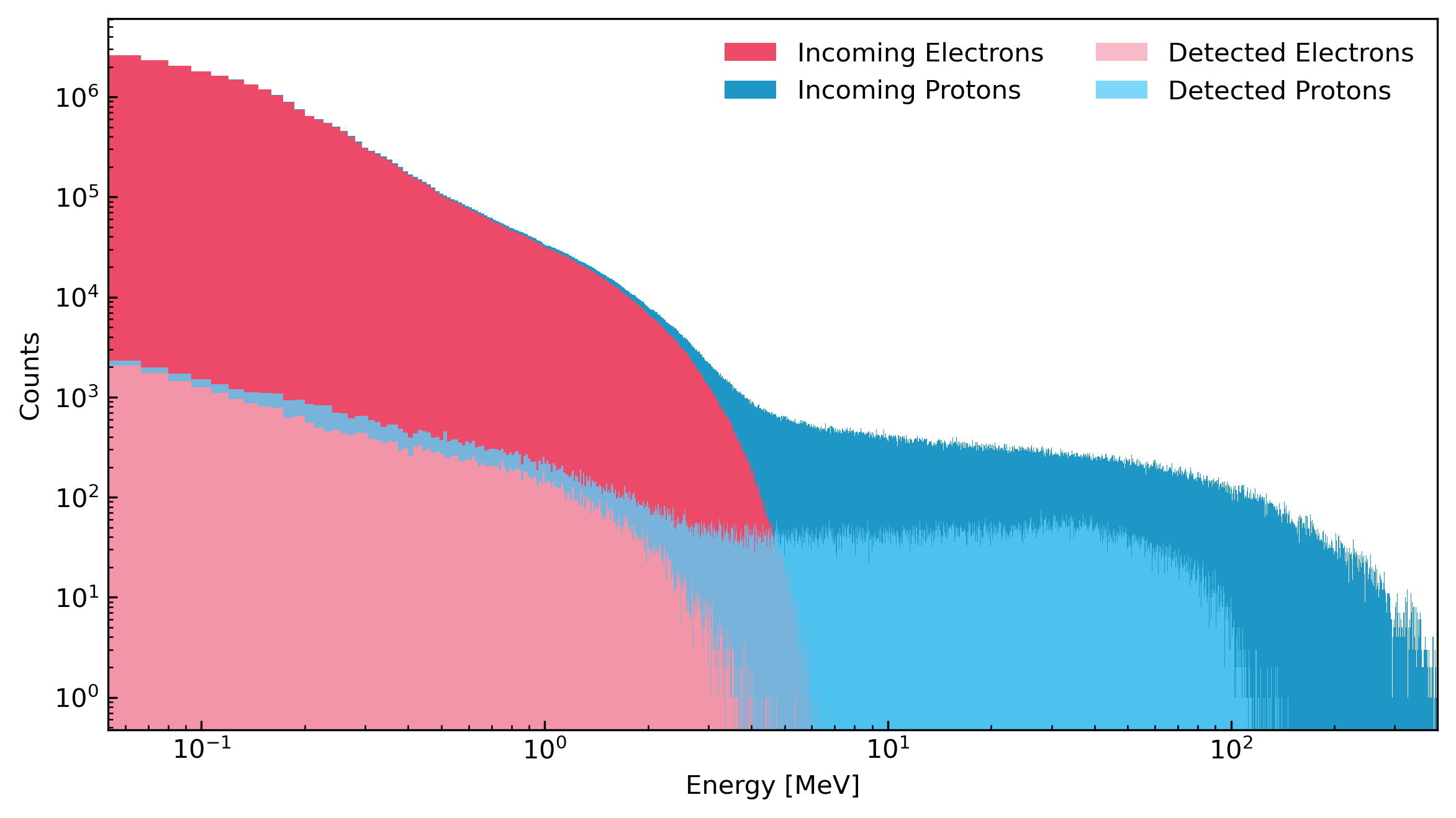}
    \caption{Simulated trapped charged particles energy spectra along the orbit of LIGHT-1. In pink and blue the incoming energy spectrum of electrons and protons is shown. In lighter colors, the energy spectrum deposited on the detector channels is shown.}
    \label{fig:geant4-spectra}
\end{figure}

\begin{figure}
    \centering
    \includegraphics[width=13cm]{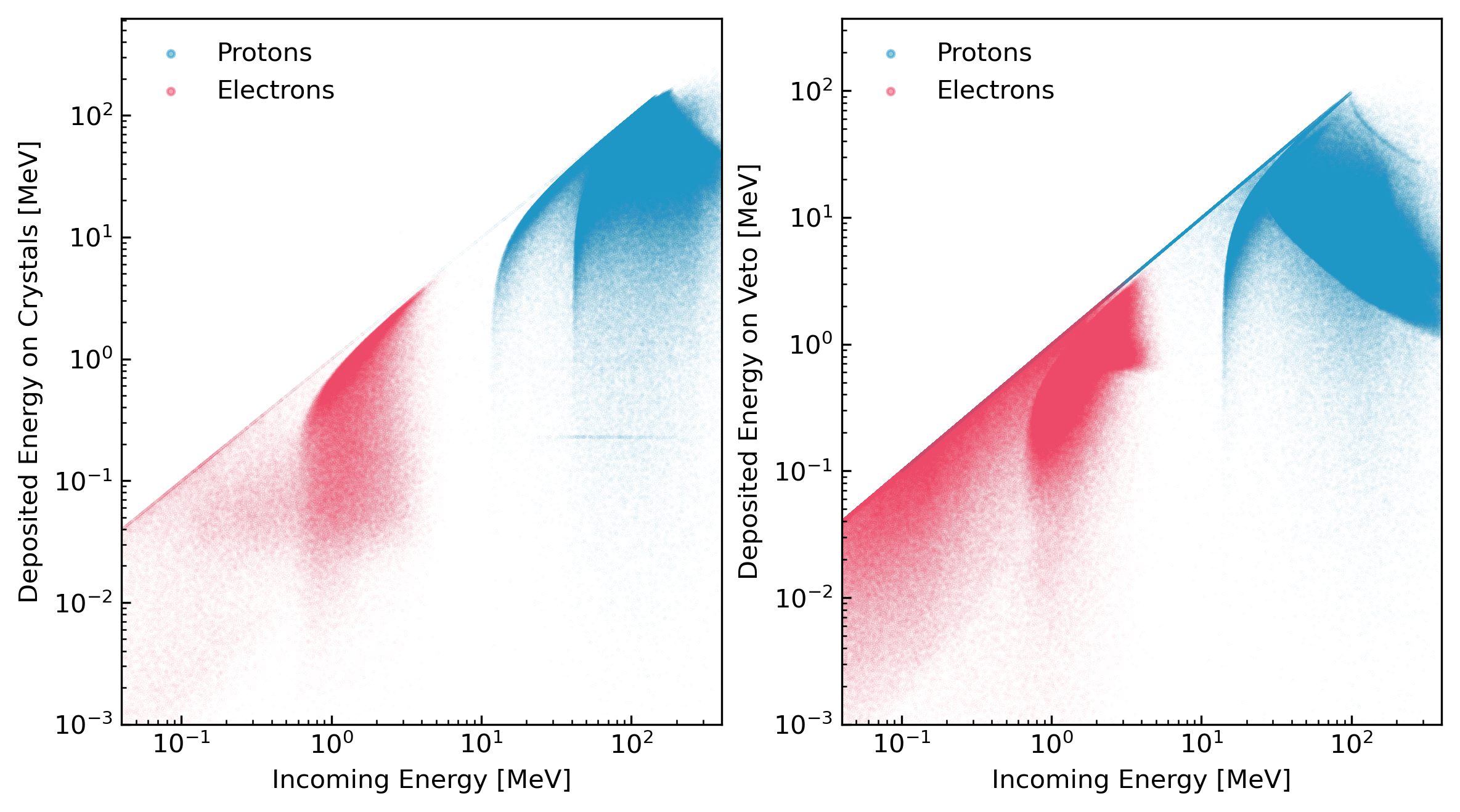}
    \caption{Energy of trapped charged particles deposited on the detector and the veto as a function of the incoming particle energy.}
    \label{fig:geant4-scatter}
\end{figure}

Using the European Space Agency's SPace ENVironment Information System (SPENVIS) \cite{SPENVIS} the flux and energy spectrum for trapped protons and electrons along the orbit of LIGHT-1 were estimated, providing the input for a Geant4-based simulation of the detectors' background spectrum. The energy deposition on the crystals vs the original particle energy spectrum is shown in Fig.~\ref{fig:geant4-spectra} while a more detailed comparison between the energy deposited on the veto vs the crystals is shown in Fig.~\ref{fig:geant4-scatter}.

%%%%%%%%%%%%%%%%%%%%%%%%%%%%%%%%%%%%%%%%%%%%%%%%%%%%%%%%%%%%%%%%%%%%%%%%%%%%%%%%%%%
%% Thermal Vacuum Tests

\subsection{Vibration Tests}
\label{subsec:Vibration-tests}

Vibration tests are typically conducted on fully assembled CubeSats to test their ability to survive launch. However, since our payload made use of  quartz windows and crystals, JAXA required additional vibration tests to be carried out on these components in isolation, prior to their integration into the main assembly.

The vibration tests were conducted  at New York University Abu Dhabi (NYUAD). A permanent magnet shaker was used to emulate the vibrational profile of the launch as well as other parameters specified by JAXA. The testing procedure is outlined below:
\begin{enumerate}
    \item A 20\,Hz - 2\,kHz sweep to identify the component's normal modes.
    \item A 7s test using a frequency sweep designed to simulate the launch conditions of a possible transport vehicle as provided by JAXA.
    \item A 20\,Hz - 2\,kHz sweep to compare the component's normal modes after exposure to launch conditions.
\end{enumerate}

This test was carried out to simulate three spacecraft candidates, namely H-II Transfer Vehicle (HTV), Cygnus NG, and SpaceX Dragon. The latter became the chosen transport vehicle for LIGHT-1. Both crystal arrays and a PMT unit were subject to these tests, all of which survived without mechanical failures, ensuring that the RAAD components were launch-safe.

\subsection{Thermal-Vacuum Tests}
\label{thermal-vacuum-tests}

\begin{figure}[t]
    \centering
    \includegraphics[width=13cm, angle=0]{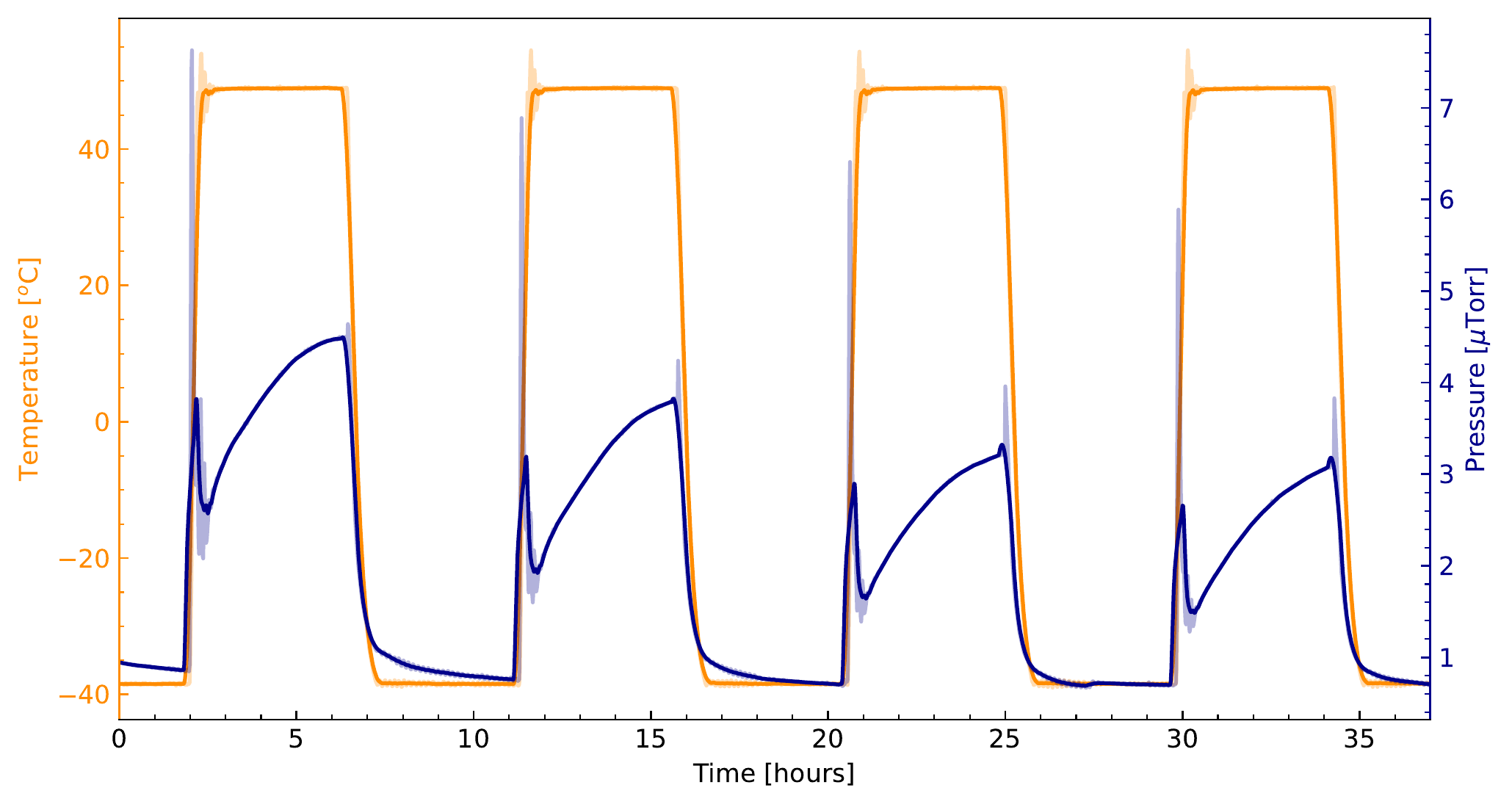}
    \caption{This figure shows the last four cycles of the thermal vacuum test conducted in Yah-Sat Lab's thermal vacuum chamber during one of the CeBr$_3$ crystal array tests.
    }
    \label{fig:TVC_crystal}
\end{figure}

Placing four crystals in the same case was a solution to the limited space problem that exists in CubeSat missions, but a concept that was not tested for space missions. Thus, it was crucial to conduct thermal vacuum tests on the crystal arrays to determine their survivability in the space environment. 
% which was never tested in space before
Two types of thermal vacuum tests were carried out on the crystal arrays:

\begin{enumerate}
    \item{One Day Tests: The arrays were left in the thermal vacuum chamber for approximately 24 hours at a set temperature of 40 $^o$C and a set pressure of 1 $\mu$Torr.}
    
    \item{Three Days Tests: The arrays were left for approximately 72 hours at a set pressure of 1 $\mu$Torr with a thermal cycle where the temperature oscillates every five hours between -40 $^o$C  and 50 $^o$C, simulating the conditions faced in space by satellites. An example of this cycle can be found in figure \ref{fig:TVC_crystal}}.
\end{enumerate}
In both thermal vacuum tests the arrays were placed on top of an insulator material, to prevent the lower side of the arrays from heating up faster than the rest of the array due to their contact with the chamber's inner metal base. These tests helped in reevaluating the structure of the arrays so that they could be optimized for space flight. In particular, a manufacturing error was uncovered that led to condensation inside one of the arrays during the thermal cycle, necessitating  its repair by the manufacturer. These tests were carried out both at YahSat Lab in Khalifa University and at NYUAD. 

%%%%%%%%%%%%%%%%%%%%%%%%%%%%%%%%%%%%%%%%%%%%%%%%%%%%%%%%%%%%%%%%%%%%%%%%%%%%%%%%%%%
%% On Flight Performance
%%%%%%%%%%%%%%%%%%%%%%%%%%%%%%%%%%%%%%%%%%%%%%%%%%%%%%%%%%%%%%%%%%%%%%%%%%%%%%%%%%%

 \section{On-Flight Perfomance}
\label{sec:on-flight-performance}

The performance and health of the payload have been evaluated continuously throughout the mission. In this section, we relate some of the experimental data observed in orbit to the parameters we have used to evaluate the operating condition of the LIGHT-1 payload.

%%%%%%%%%%%%%%%%%%%%%%%%%%%%%%%%%%%%%%%%%%%%%%%%%%%%%%%%%%%%%%%%%%%%%%%%%%%%%%%%%%%
%% Instrument Health
\subsection{Instrument Health}
\label{subsec:instrument-health}

\begin{figure}[t]
    \centering
    \includegraphics[width=13cm, angle=0]{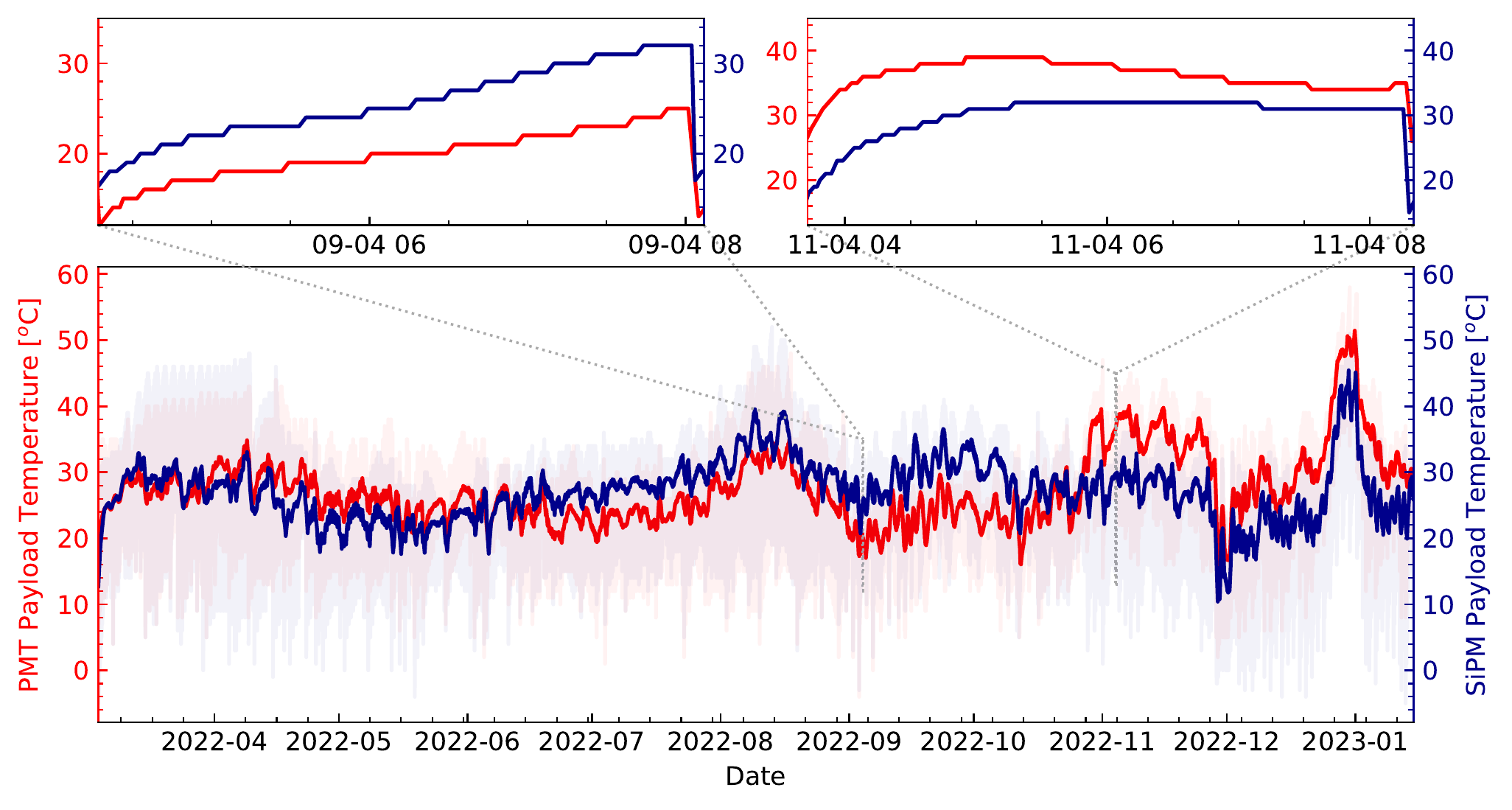}
    \caption{Temperature of the PMT and MPPC Payloads from the Payload Commissioning (2022-04) until the de-orbit (2023-01) of LIGHT-1. The temperatures of the PMT and MPPC detectors are shown in red and blue respectively (after smoothing by a Savitzky-Golay with a window length of 1 day). The unfiltered temperature is overlaid with lighter colors. Using two insets we highlight the temperature increase during two different duty cycles. As can be seen from the two insets of randomly picked duty cycles, the temperature variation is high and difficult to predict as it varies from orbit to orbit.}
    \label{fig:payload-temp}
\end{figure}

Throughout the mission, we have monitored the health of the hardware through onboard sensors. In particular, the Texas Instruments LM71 temperature sensor embedded in each of the readout electronics boards measured the payload temperature over the lifespan of the mission. The resulting measurements from the commissioning of the payload until de-orbit are shown in Fig.~\ref{fig:payload-temp}.

The payload's temperature data during orbit is important to correct for detector gain drift  \cite{DiGiovanni2019}, and for adjusting the operations duty cycle in flight to avoid overheating. As can be seen in the insets of Fig.~\ref{fig:payload-temp}, which represent the temperature recorded while the payload was operational, during a single orbit the temperature rises significantly when the payload is operating. This occurs due to the power dissipated by the electronics. Already in its design phase, the payload mission concept has been tailored around a reduced 50 $\% $  duty cycle in order to meet the mission requirements in terms of the power budget. From the commissioning to the de-orbit phase, the electronics have been subjected to 1861 power cycles. Under these conditions and for the entire mission, no emergency procedure has been triggered due to temperature failure.

\paragraph{Pulse Per Second Signal Loss.}
\label{par:pps-loss}

 By design, a key feature of the RAAD electronics is the ability of assigning a sub-microsecond time stamping to each acquired event. To be compliant with the power constraints of the mission, since the design phase, the use of fast electronics capable of coping with the typical time response of MPPCs and PMTs (order of $\mathrm{\sim 1~GS/s}$) could not be considered. 
 
 The solution implemented on the LIGHT-1 timing architecture utilizes a distributed disciplined clock, controlled by a Phase Locked Loop (PLL). 
 
 The PLL generates an high frequency output ($\mathrm{10~MHz}$) from a low frequency reference ($\mathrm{1~Hz}$). The RAAD timing circuit concept is based on a PLL using the $\mathrm{1~Hz}$ Pulse-Per-Second (PPS) hardware signal obtained from the spacecraft GPS receiver coupled to a $\mathrm{10~ MHz}$ oscillator.

However, an unidentified failure  made the PPS signal unavailable. As a consequence, the capability to calculate and assign the timestamp to events on the payload data buffers was compromised from the beginning of the LEOP Phase until the Science Run began (see Fig.~\ref{fig:light1life}). 

 To overcome this issue, a software patch has been implemented during the Payload Commissioning phase (see Fig.~\ref{fig:light1life}). The patch uses the onboard computer's signal to provide a, less accurate, PPS signal digitally. We used further reconstruction techniques on the ground after obtaining the data in order to retrieve the original accuracy. 

\paragraph{Seasonal Temperature Variation.}
\label{par:seasonal-temperature-variation}

\begin{figure}
    \centering
    \includegraphics[width=13cm, angle=0]{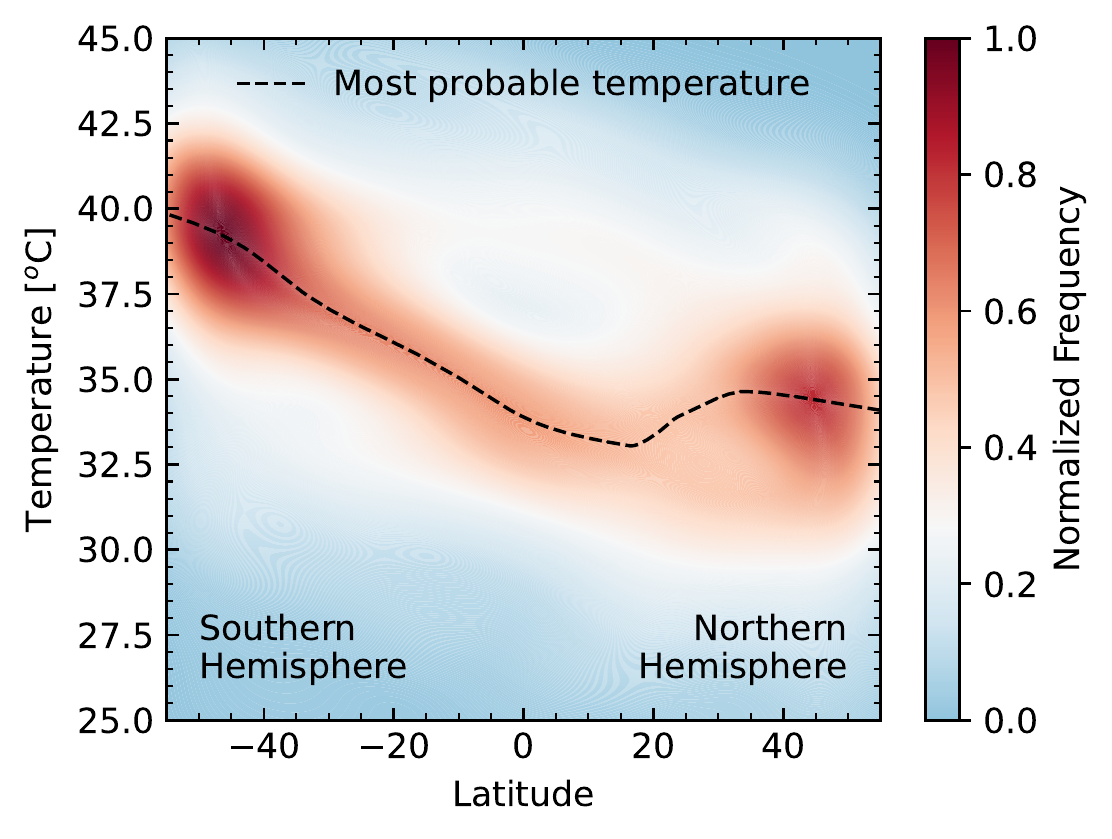}
    \caption{Temperature of the PMT payload as a function of latitude during 1$^\text{st}$ - 23$^\text{rd}$ of October 2022. The shaded region shows the probability density of the temperature over the latitude, while the red, dotted line is the maximum of this distribution. We observe the characteristic hotter temperatures for the southern hemisphere during October.}
    \label{fig:latitude-temp}
\end{figure}

We further verified the health of the onboard electronics by tracking the temperature of the satellite during October as a function of latitude. In Fig.~\ref{fig:latitude-temp} the dotted line shows the most probable temperature measured by the satellite as a function of latitude. In the plot, we observe the expected seasonal variation, namely hot temperatures in the southern hemisphere due to increased sunlight during October, and, conversely, colder temperatures in the northern hemisphere. Such results were used to verify the integrity of our reconstructed data.

%%%%%%%%%%%%%%%%%%%%%%%%%%%%%%%%%%%%%%%%%%%%%%%%%%%%%%%%%%%%%%%%%%%%%%%%%%%%%%%%%%%
%% Conclusion
%%%%%%%%%%%%%%%%%%%%%%%%%%%%%%%%%%%%%%%%%%%%%%%%%%%%%%%%%%%%%%%%%%%%%%%%%%%%%%%%%%%

\section{Conclusions}
\label{sec:conclusion}

In this paper, we describe RAAD, the scientific payload of the LIGHT-1 CubeSat mission, designed to detect fast (< 1ms) X and gnmma-ray transients with two detectors fitting in  1.7 CubeSat units. The payload underwent rigorous physical testing and simulation in order to space-quality the instrument. In particular, the thermal vacuum and vibration tests of the individual components and the assembly certified RAAD according to JAXAs specifications for surviving typical launch conditions for such instrumentation. Space qualification was further validated by carrying out a modal analysis of the apparatus, showing vibrational modes within the approved range for withstanding launch conditions.
The expected trapped charged particle flux along the orbit of the detector was estimated using data from the European Space Agency's SPace ENVironment Information System (SPENVIS) \cite{SPENVIS} allowing us to carry out an exhaustive Geant4 particle physics simulation to predict the expected background  and optimize the detection threshold accordingly as well as the geometry and optical insulation of the instrument.

The LIGHT-1 satellite was launched from the Kennedy Space Center on December 21st, 2021, on a SpaceX rocket, which docked at the International Space Station the day after. It was deployed on February 3rd, 2022. Contact was lost on January 14, 2023.  While the analysis of the data is ongoing, the housekeeping data presented show that, with the exception of the failure that led to the PPS loss, RAAD withstood the launch stress as expected, and the detectors operated throughout the mission.

%%%%%%%%%%%%%%%%%%%%%%%%%%%%%%%%%%%%%%%%%%%%%%%%%%%%%%%%%%%%%%%%%%%%%%%%%%%%%%%%%%%
%% Appendices\textbf{}
%%%%%%%%%%%%%%%%%%%%%%%%%%%%%%%%%%%%%%%%%%%%%%%%%%%%%%%%%%%%%%%%%%%%%%%%%%%%%%%%%%%

%%%%%%%%%%%%%%%%%%%%%%%%%%%%%%%%%%%%%%%%%%%%%%%%%%%%%%%%%%%%%%%%%%%%%%%%%%%%%%%%%%%
%% Aknowledgments
%%%%%%%%%%%%%%%%%%%%%%%%%%%%%%%%%%%%%%%%%%%%%%%%%%%%%%%%%%%%%%%%%%%%%%%%%%%%%%%%%%%

\acknowledgments

We gratefully acknowledge the support of the UAE Space Agency through the 2018 MiniSat Competition, and the NYUAD Kawader program for supporting one of the authors (L. AlKindi). Special thanks to Sebastien Celestin for providing TGF models.  We also thank Khalifa University and NSSA for funding their master's students to work on the CubeSat's Bus system design. Finally, we express our gratitude to the NYUAD Core Technology Platforms for their invaluable assistance and particularly to the machine shop team for the realization of the aluminum enclosures of the detection targets.

%%%%%%%%%%%%%%%%%%%%%%%%%%%%%%%%%%%%%%%%%%%%%%%%%%%%%%%%%%%%%%%%%%%%%%%%%%%%%%%%%%%
%% Reference
%%%%%%%%%%%%%%%%%%%%%%%%%%%%%%%%%%%%%%%%%%%%%%%%%%%%%%%%%%%%%%%%%%%%%%%%%%%%%%%%%%%

\bibliographystyle{JHEP}
\bibliography{biblio.bib}

\end{document}